\documentstyle[preprint,aps]{revtex}
\input{epsf}

\begin{document}
\draft
\title{Plastic flow of persistent currents in two dimensional strongly 
interacting systems}

\author
{Richard Berkovits and Yshai Avishai$^{\dag}$}

\address{
The Minerva Center for the Physics of Mesoscopics, Fractals and Neural 
Networks,\\ Department of Physics, Bar-Ilan University,
Ramat-Gan 52900, Israel}

\address{
$\dag$also at Department of Physics, Ben-Gurion University, Beer-Sheva, Israel}

\date{\today}
\maketitle

\begin{abstract}
The local persistent current in two dimensional strongly 
interacting systems is investigated. As the interaction strength is enhanced
the current in the sample undergoes a transition from diffusive to ordered
flow. The strong interacting flow has the characteristics of a plastic flow
through dislocations in the pinned charge density wave which develops in the
system at low densities.
\end{abstract}


\narrowtext

\newpage

The behavior of interacting electrons
in random potentials has drawn much interest because of its 
relevance to many different phenomena such as 
persistent currents \cite{gmw},
the two-dimensional MIT (metal-insulator transition) \cite{2dmit},
and the charging spectra of quantum dots \cite{sba,shw,cm}.
It is generally believed that the large
amplitudes of persistent currents measured for mesoscopic rings in the
diffusive regime \cite{exp1,exp2} are the result of the suppression of the
influence of disorder by interactions, although no quantitative calculation
for a realistic model of a 3D metallic system has been forthcoming. 

Much recent work has concentrated on the influence of interactions on 2DEG
(2D Electron Gas). It is speculated that the 
experimentally observed 2D MIT\cite{2dmit} 
is due to e-e (electron-electron) interactions\cite{2dmitt},
while  indications of the influence of e-e interactions in the addition 
spectrum of disordered
quantum dots are mounting \cite{sba,shw}. For typical 2DEG devices
the density is rather low ($n<3\times10^{11} {\rm cm}^{-1}$) which corresponds
to a ratio between the the Coulomb and Fermi energy $r_s=e^2/v_f>1$ 
(where $v_f$ is the Fermi velocity) \cite{ba}.
In this region the simple RPA (random phase approximation) no longer holds,
and correlations play an important role \cite{ba}. 

Persistent currents for a clean 2DEG ring has been recently measured 
and shown to be of the expected magnitude\cite{exp3}. 
No measurements have yet been 
performed in the diffusive regime, but numerical calculations 
indicate that the persistent current will be significantly 
enhanced (compared to the non-interacting value) by the e-e interactions
even for spinless electrons \cite{pc}.

In this paper we would like to clarify the mechanism of this enhancement in
2DEG rings. A useful clue may be obtained from the study of the distribution
of the current in different realizations of disorder \cite{dis}. While in
the non-interacting case the distribution is broad with almost the same
probability for dia or paramagnetic currents
leading to small paramagnetic
average currents, in the interacting case
the distribution is predominately paramagnetic leading to large paramagnetic 
average currents. In this paper we shall show that the local persistent
current undergoes a transition from a diffusive behavior for the 
non-interacting regime to a strongly correlated plastic flow through a small
number of channels for stronger e-e interactions. This leads to a change in the
local distribution of the current and to the enhancement of the total
average current.

Thus, there are two distinct mechanisms for the enhancement of the persistent 
current in 2DEG systems. One which is relevant to the high density weak 
interaction regime which is described analytically by the diagrammatic 
perturbation treatment \cite{ae,sc}, or by the Hartree-Fock picture 
\cite{mon}, and treated numerically by various Hartree-Fock approximations 
\cite{rrb,ky,bp,vk}. The other
is the low density strongly interacting regime which is characterized
by the appearance of correlations in the electronic density and plastic
flow of the persistent current.

The local persistent current of an interacting 2D cylinder of
circumference $L_x$ and height $L_y$ threaded by a flux $\Phi$ was calculated.
This system is known to show a 
large enhancement of the persistent current in the diffusive regime \cite{pc}.
The Hamiltonian is given by:
\begin{eqnarray}
H= \sum_{k,j} \epsilon_{k,j} a_{k,j}^{\dag} a_{k,j} -
i \sum_{k,j} J_{k,j}^x +  J_{k,j}^y +
U  \sum_{k,j>l,p} {{a_{k,j}^{\dag} a_{k,j}
a_{l,p}^{\dag} a_{l,p}} \over 
{|\vec r_{k,j} - \vec r_{l,p}|}}
\label{hamil}
\end{eqnarray}
where
\begin{eqnarray}
J_{k,j}^x &=& - i V 
\exp(i\Phi s/L_x) a_{k,j+1}^{\dag} a_{k,j} - {\rm h.c}, \nonumber \\
J_{k,j}^y &=& - i V 
a_{k+1,j}^{\dag} a_{k,j} - {\rm h.c},
\label{jx}
\end{eqnarray}
and  $a_{k,j}^{\dag}$
is the fermionic creation operator,
$\epsilon_{k,j}$ is the energy of a site ($k,j$), which is chosen 
randomly between $-W/2$ and $W/2$ with uniform probability, $V$
is a constant hopping matrix element and
$s$ is the lattice constant. 
The distance $|\vec r_{k,j} - \vec r_{l,p}|=
(\min\{(k-l)^2,(L_x/s - (k-l))^2\}
+\min\{(j-p)^2,(L_y/s-(j-p))^2\})^{1/2}.$
The interaction term represents a Coulomb
interaction between electrons confined to a 2D
cylinder embedded in a 3D space with $U = e^2/s$,

We consider a $4 \times 4$ lattice with $m=16$ sites
and $n=8$ electrons. The many-particle Hamiltonian then may be represented
by a $12870 \times 12870$ matrix, which is exactly diagonalized. 
The many-particle ground state $|\Psi(\Phi)\rangle$ for $\Phi=\pi/2$
is calculated for $500$ different realizations of disorder
in the diffusive regime ($W=8V$ \cite{pc}) for several values of 
interaction $U$. The local persistent current
\begin{eqnarray}
I_{k,j}^{a}(\Phi) = \langle \Psi(\Phi)|J_{k,j}^a|\Psi(\Phi) \rangle,
\label{ix}
\end{eqnarray}
(where $a= x,y$) is calculated for each realization.

\begin{figure}
\leftline{\ \ \ \ \ \ \ \ \ \ \ \ \ \ 
\ \ U=0 
\ \ \ \ \ \ \ \ \ \ \ \ \ \ \ \ \ \ \ \ \ \ \ \ \ \ \ \ \ \ \ \ \ \ \ 
\ \ \ \ \ \ \ \ \ 
\ \ \ \ \ \ \ \ \ \ \ \ \ \ \ \ \ \ \ \ U=10V}
\vspace{-1.2in}
\leftline{\epsfxsize = 2.5in \epsffile{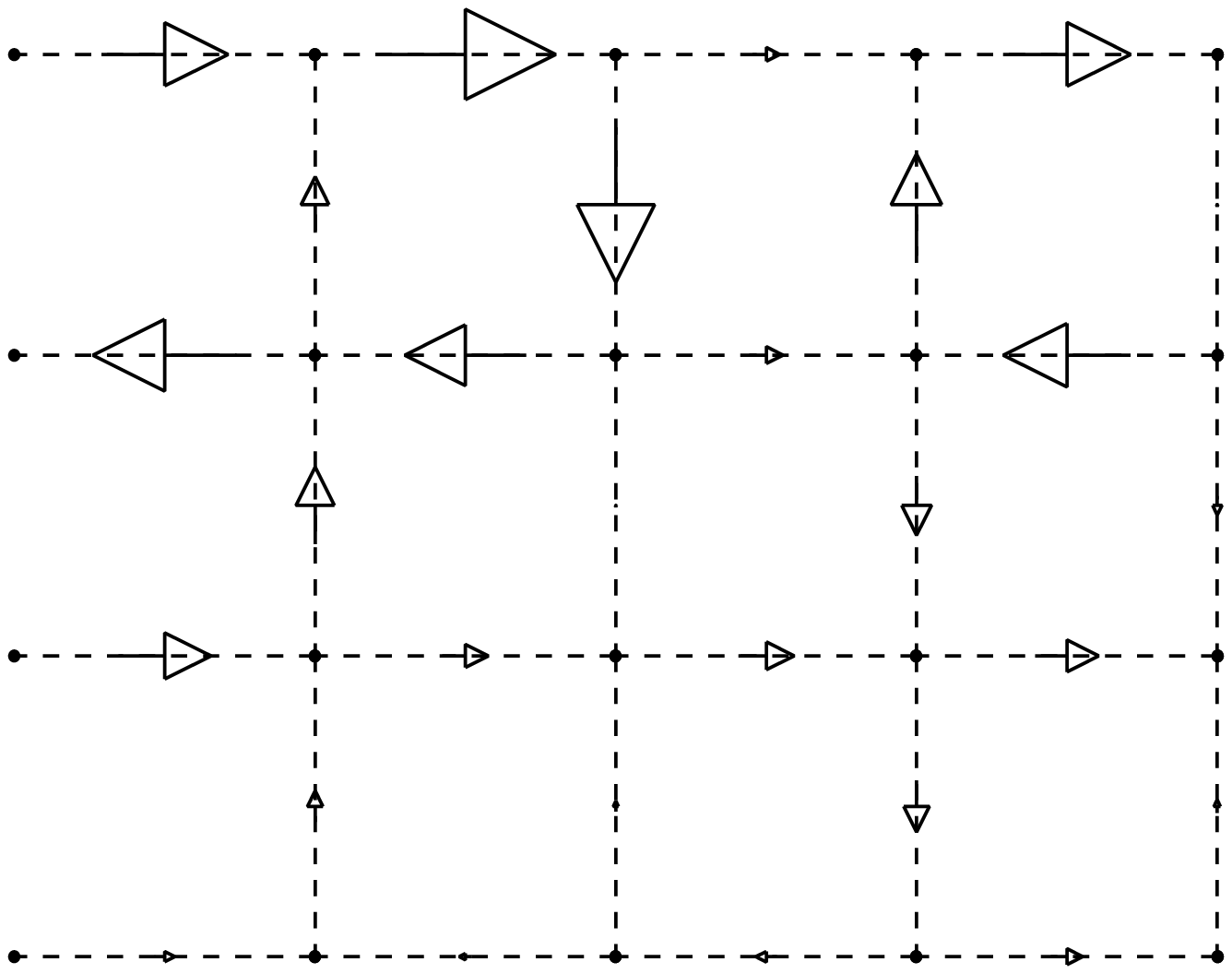}}
\vspace{-3.45in}
\rightline{\epsfxsize = 2.5in \epsffile{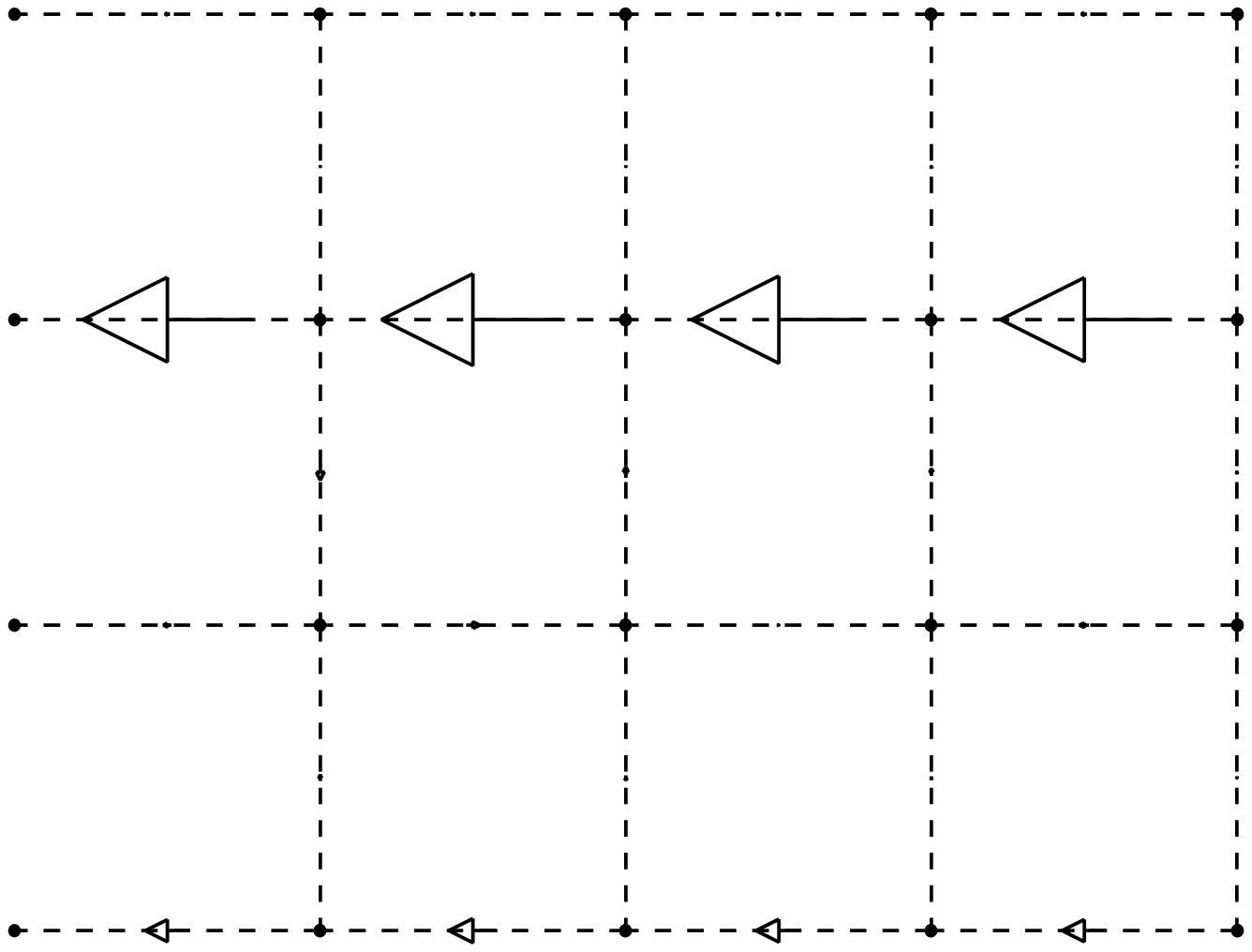}}
\vspace{-1.5in}
\leftline{\epsfxsize = 2.5in \epsffile{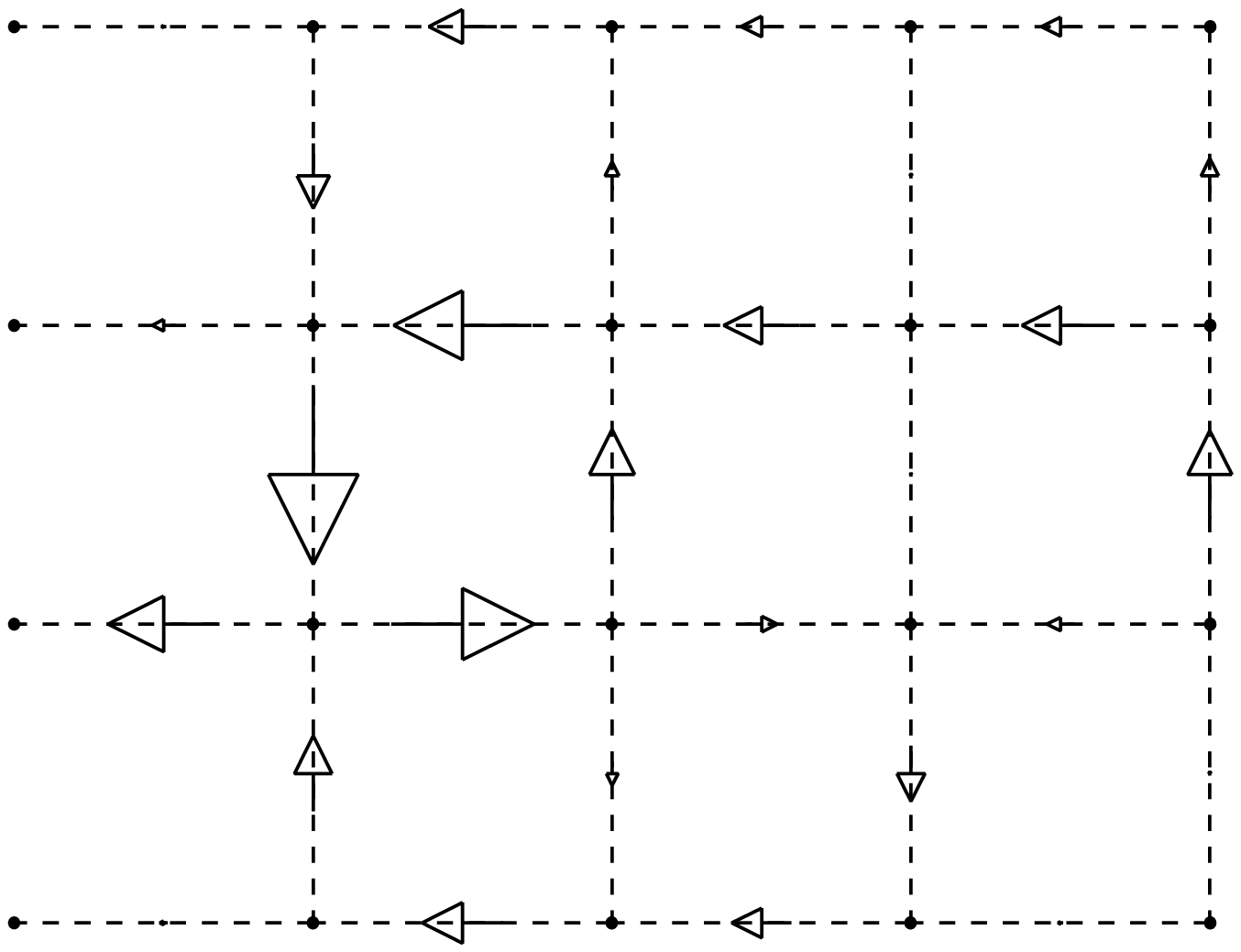}}
\vspace{-3.45in}
\rightline{\epsfxsize = 2.5in \epsffile{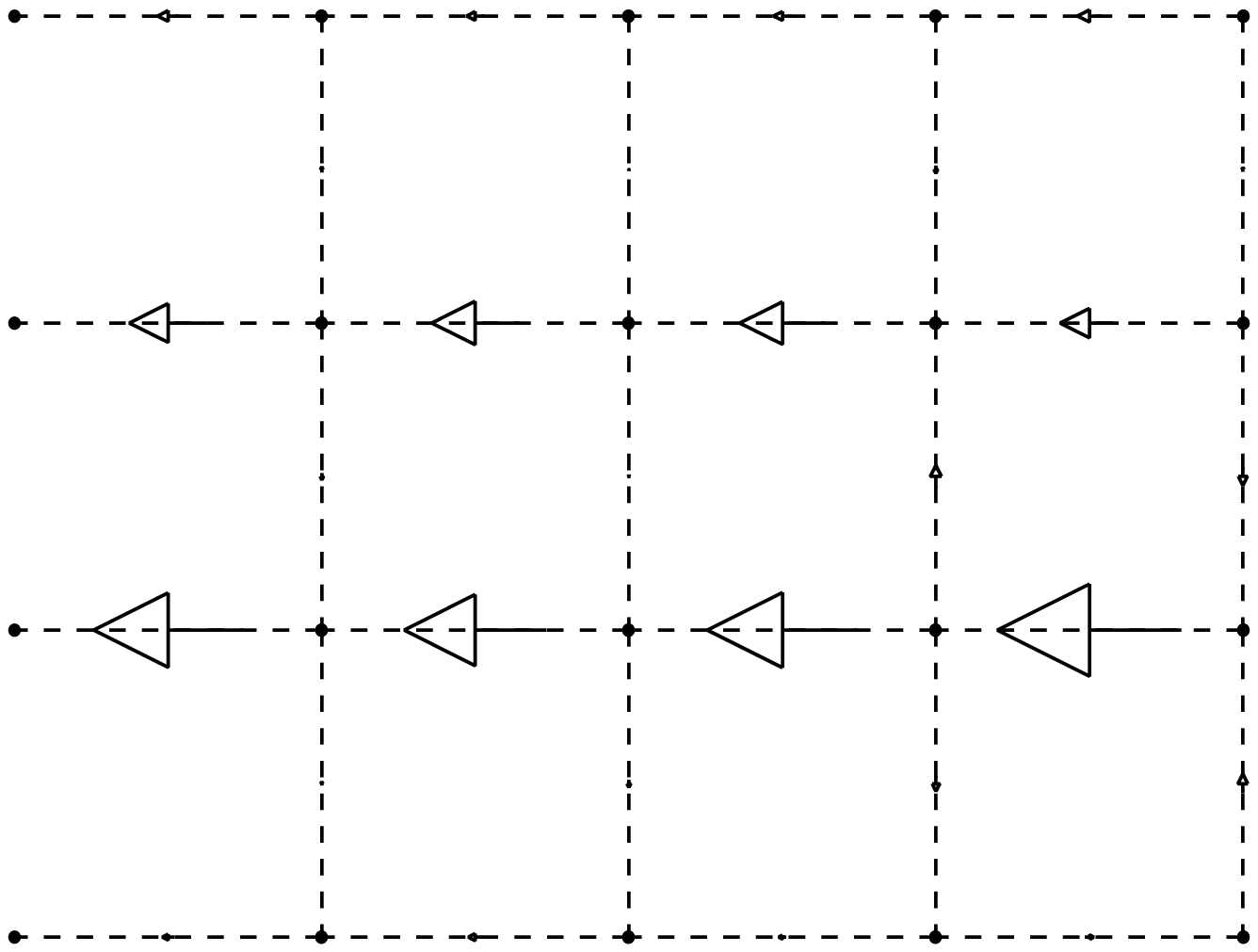}}
\vspace{-1.5in}
\leftline{\epsfxsize = 2.5in \epsffile{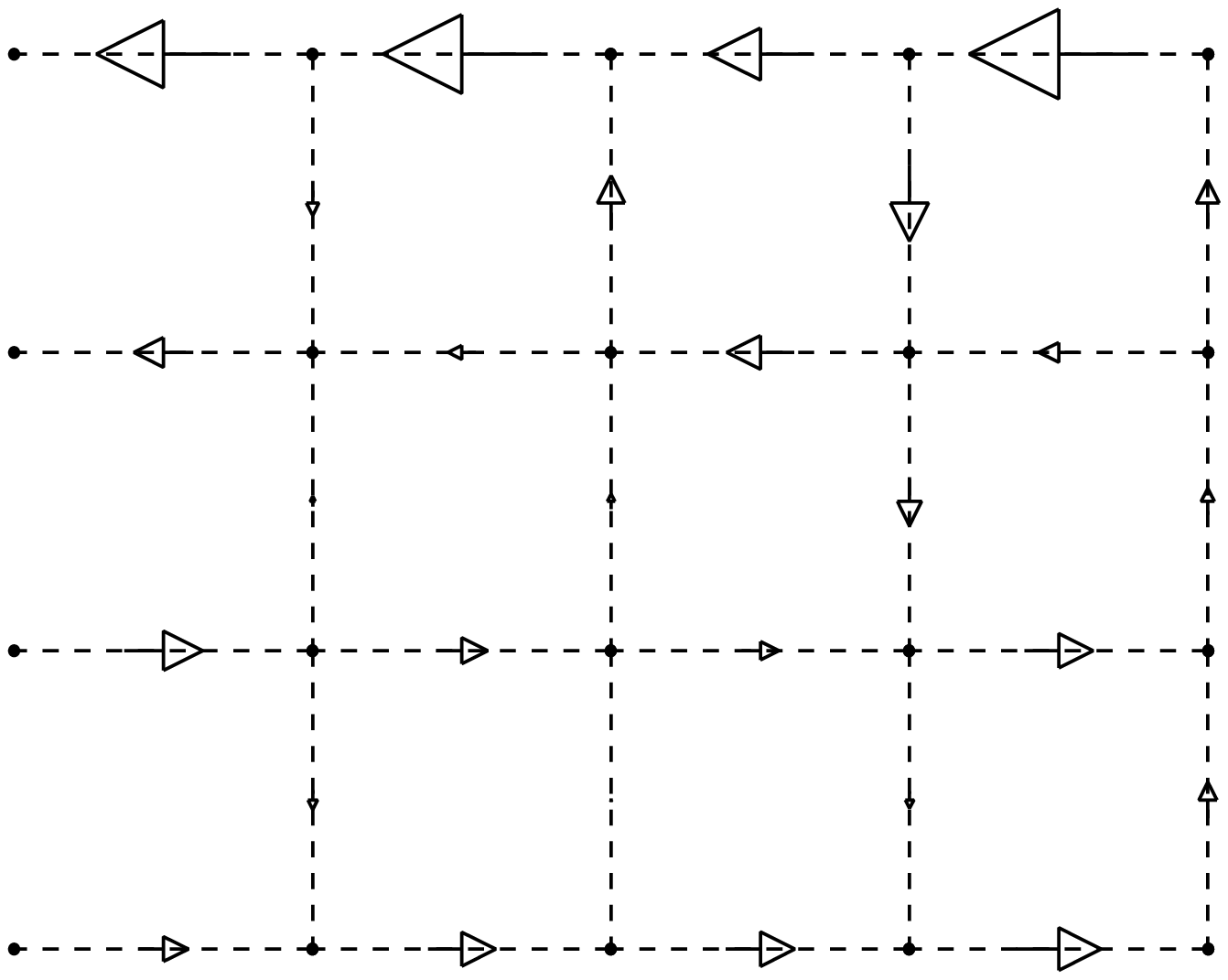}}
\vspace{-3.45in}
\rightline{\epsfxsize = 2.5in \epsffile{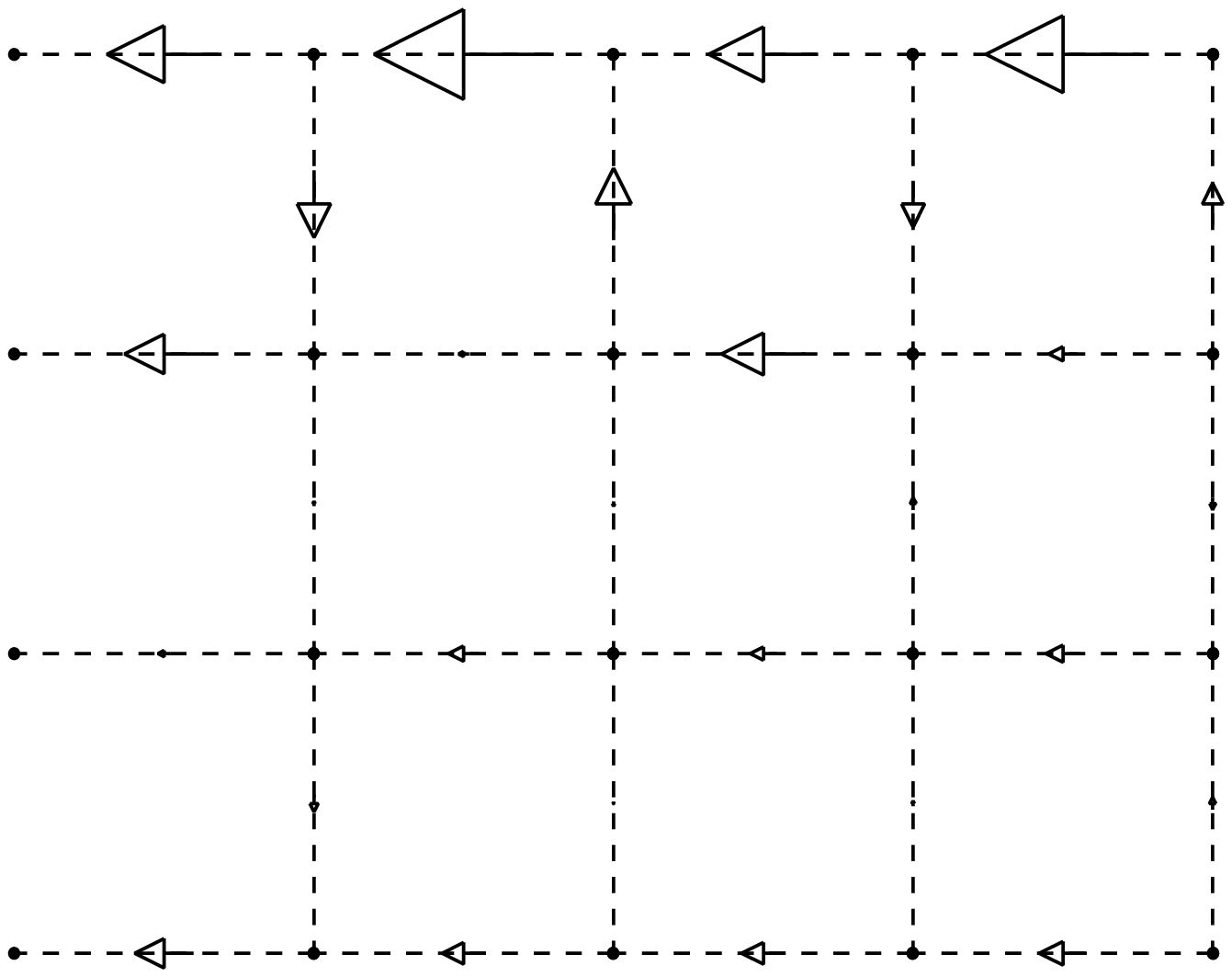}}
\caption {The local current for several realizations for different values
of $U$. Left column: non interacting case: right column: the same realization
with $U=10V$. The size of the arrows are proportional to the magnitude of the
current
\label{fig.1}}
\end{figure}

Local currents in typical realizations are plotted in Fig. \ref{fig.1}. 
It can be seen that the
current for the non-interacting case $U=0$ is diffusive, the current flows
in all directions, there are current loops and there is no obvious
long range correlations. On the other hand for $U=10V$, which corresponds to
$r_s\sim\sqrt{\pi/2}(U/4V)\sim 3$ a value which could be easily obtained
in contemporary 2DEG devices, the behavior is totally different. The persistent
current flows through a few channels in the sample, there are no currents 
flowing in the opposite  directions, no closed loops, and obvious
correlations. This situation reminds us of plastic flow in lattices,
for example the flow of magnetic vertexes in driven Abrikosov lattice pinned 
by random pinning centers \cite{nori}.

\begin{figure}
\leftline{.}
\vspace{-1.2in}
\leftline{\epsfxsize = 2.5in \epsffile{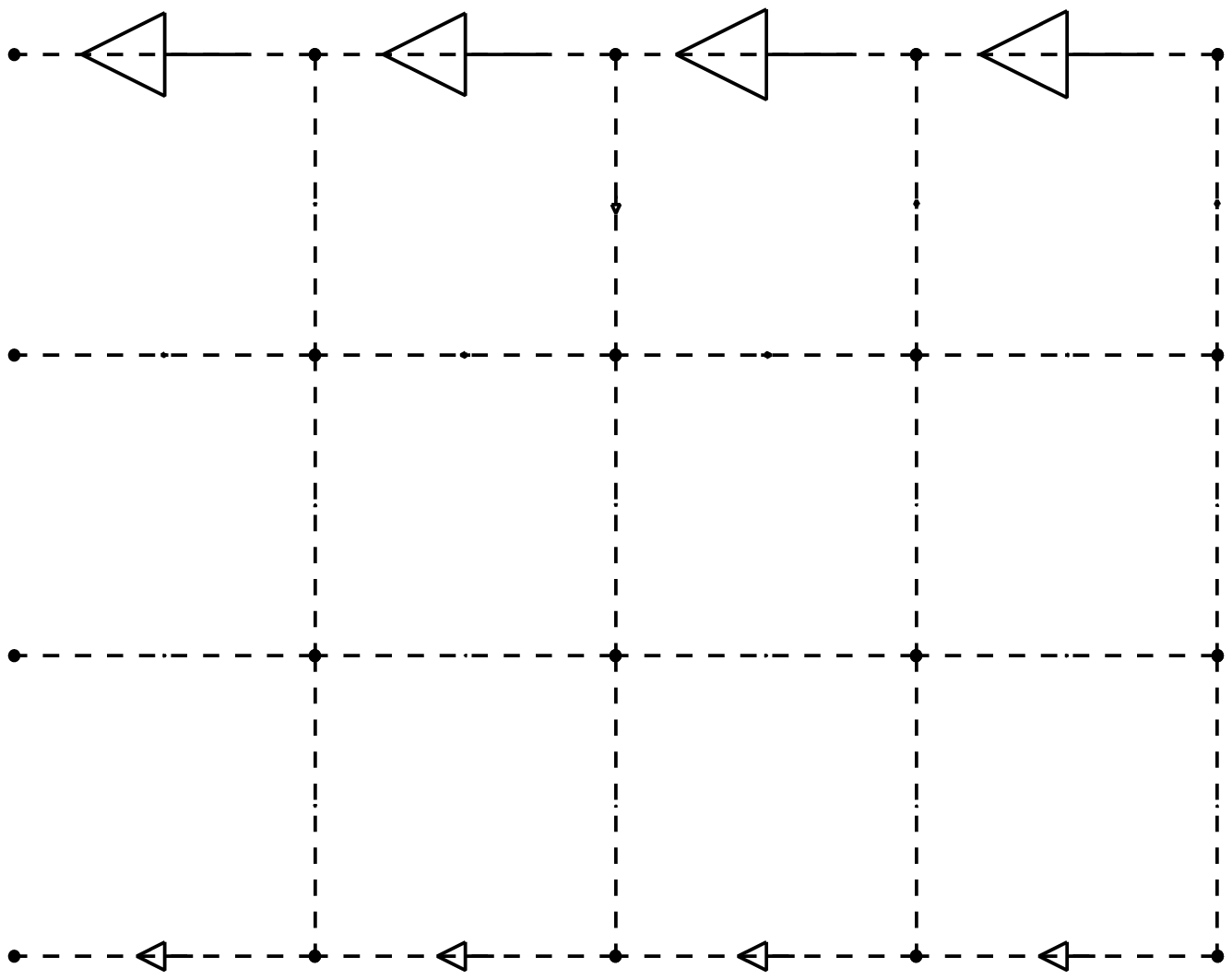}}
\vspace{-3.2in}
\rightline{\epsfxsize = 2.5in \epsffile{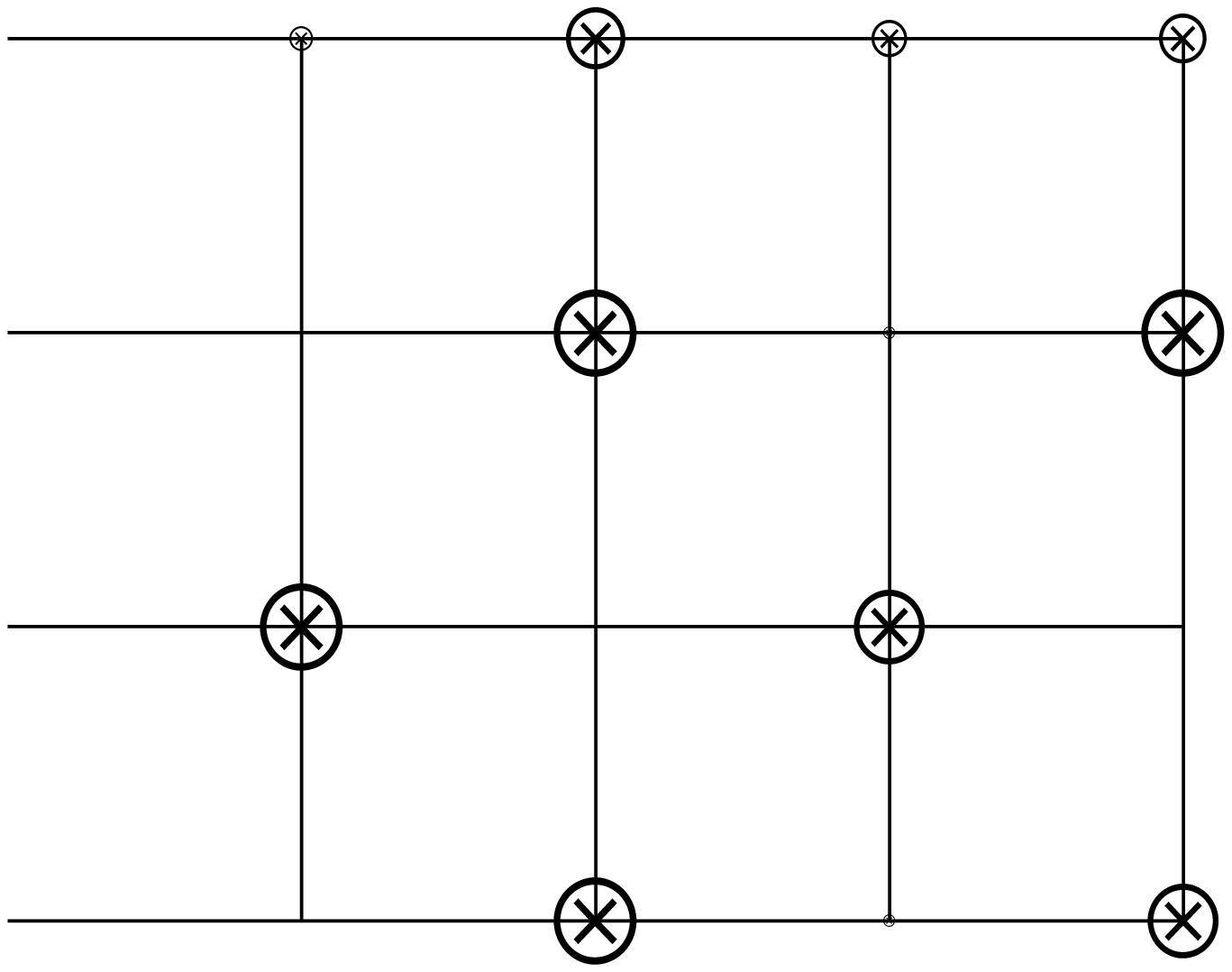}}
\vspace{-1.5in}
\leftline{\epsfxsize = 2.5in \epsffile{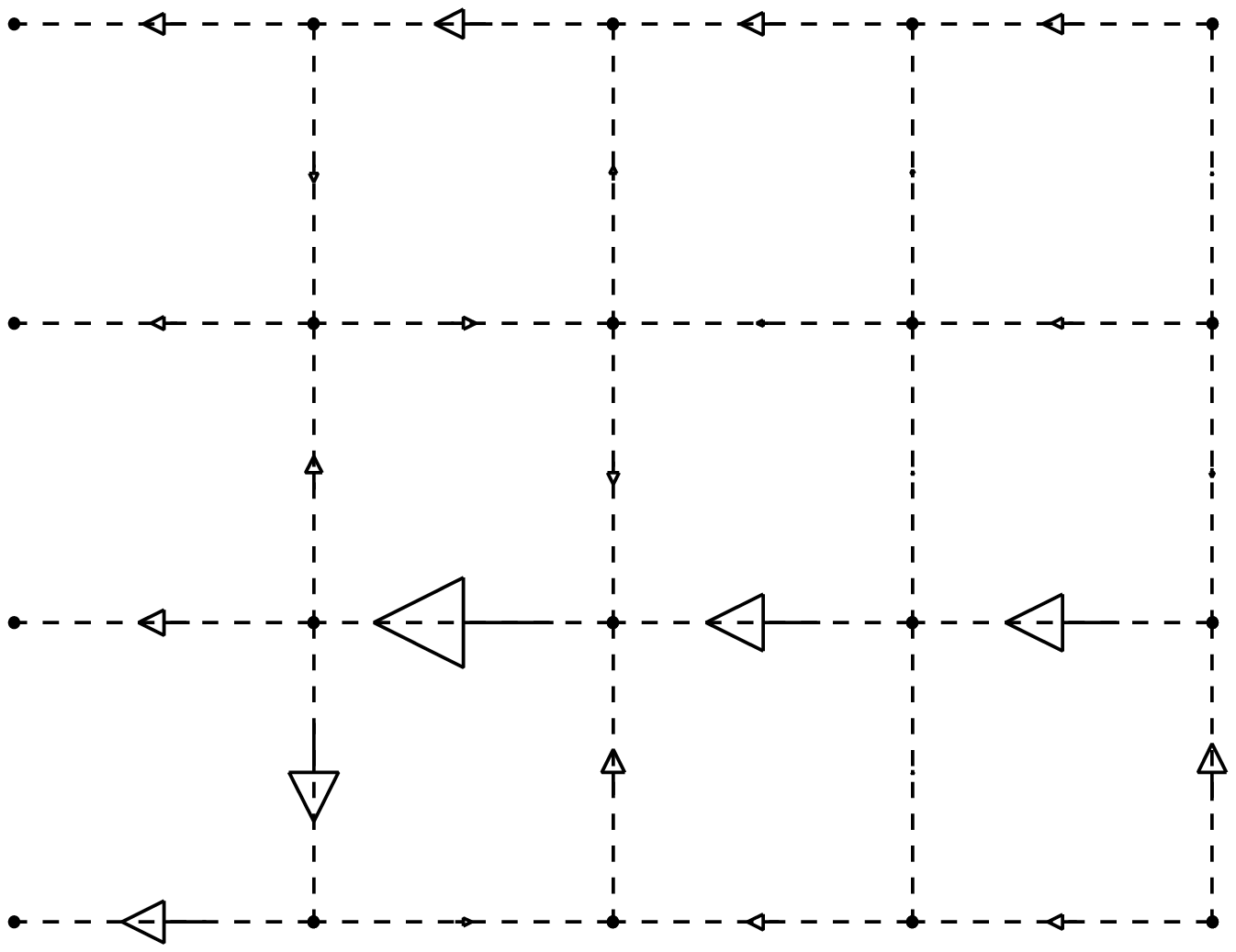}}
\vspace{-3.2in}
\rightline{\epsfxsize = 2.5in \epsffile{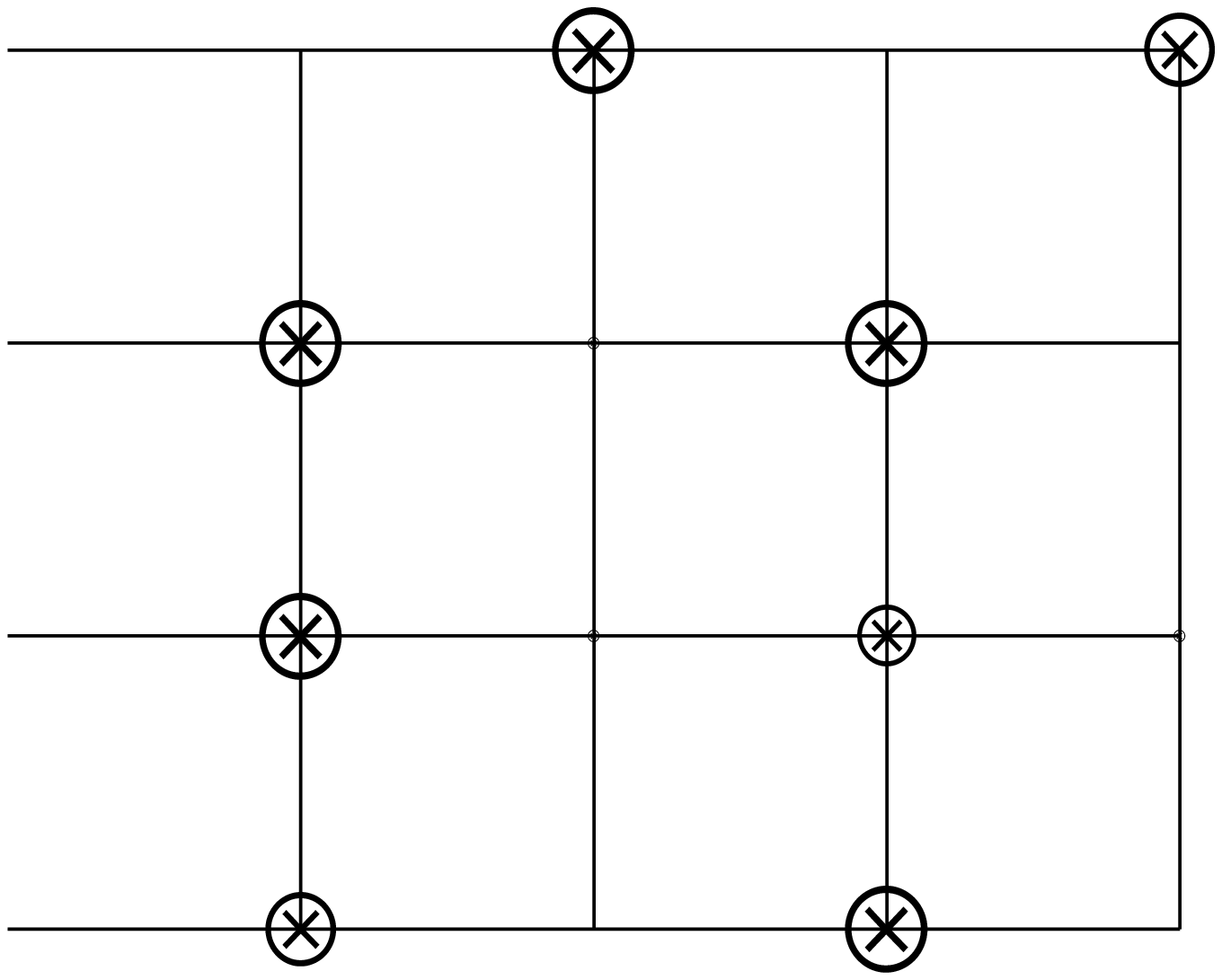}}
\vspace{-1.5in}
\leftline{\epsfxsize = 2.5in \epsffile{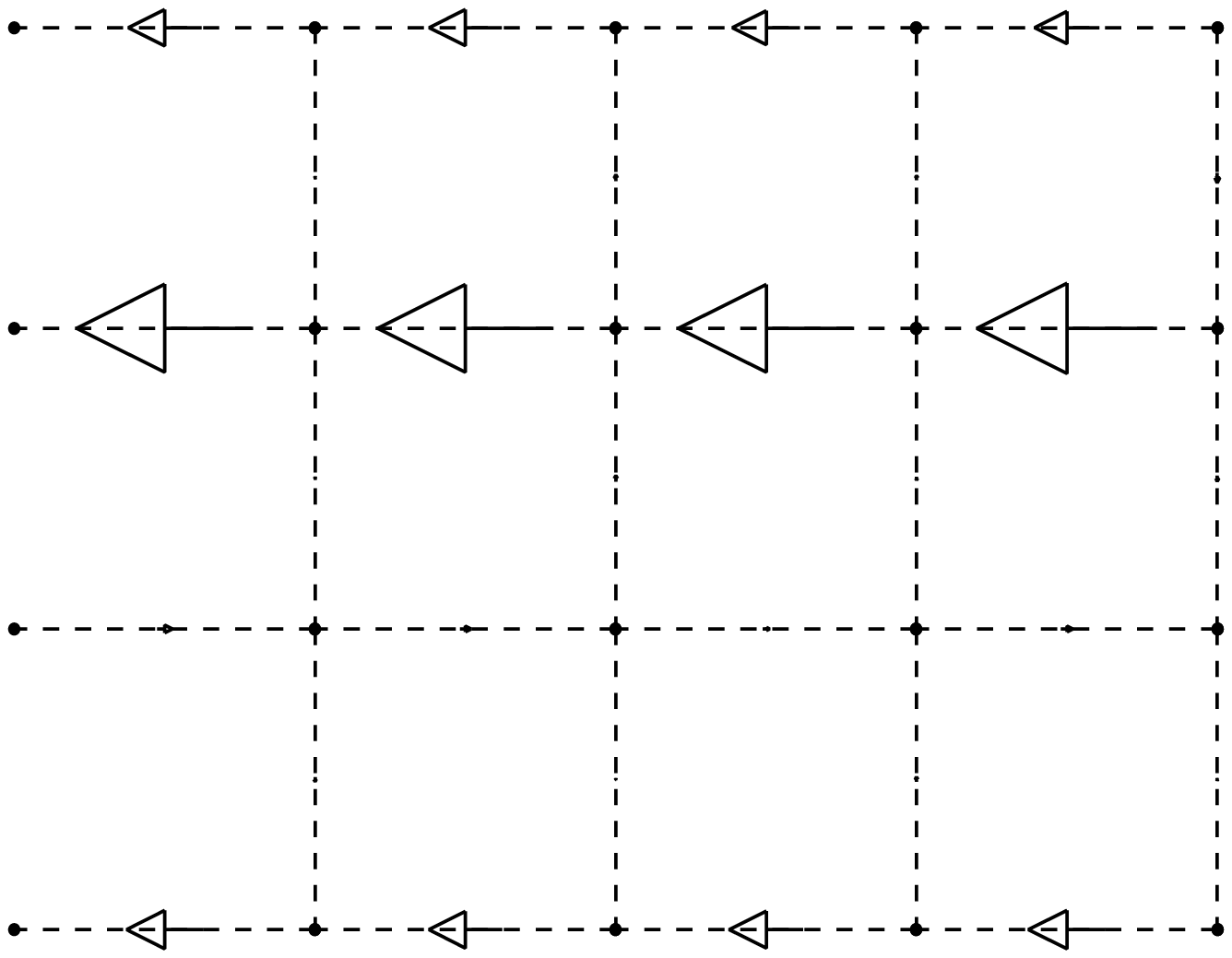}}
\vspace{-3.2in}
\rightline{\epsfxsize = 2.5in \epsffile{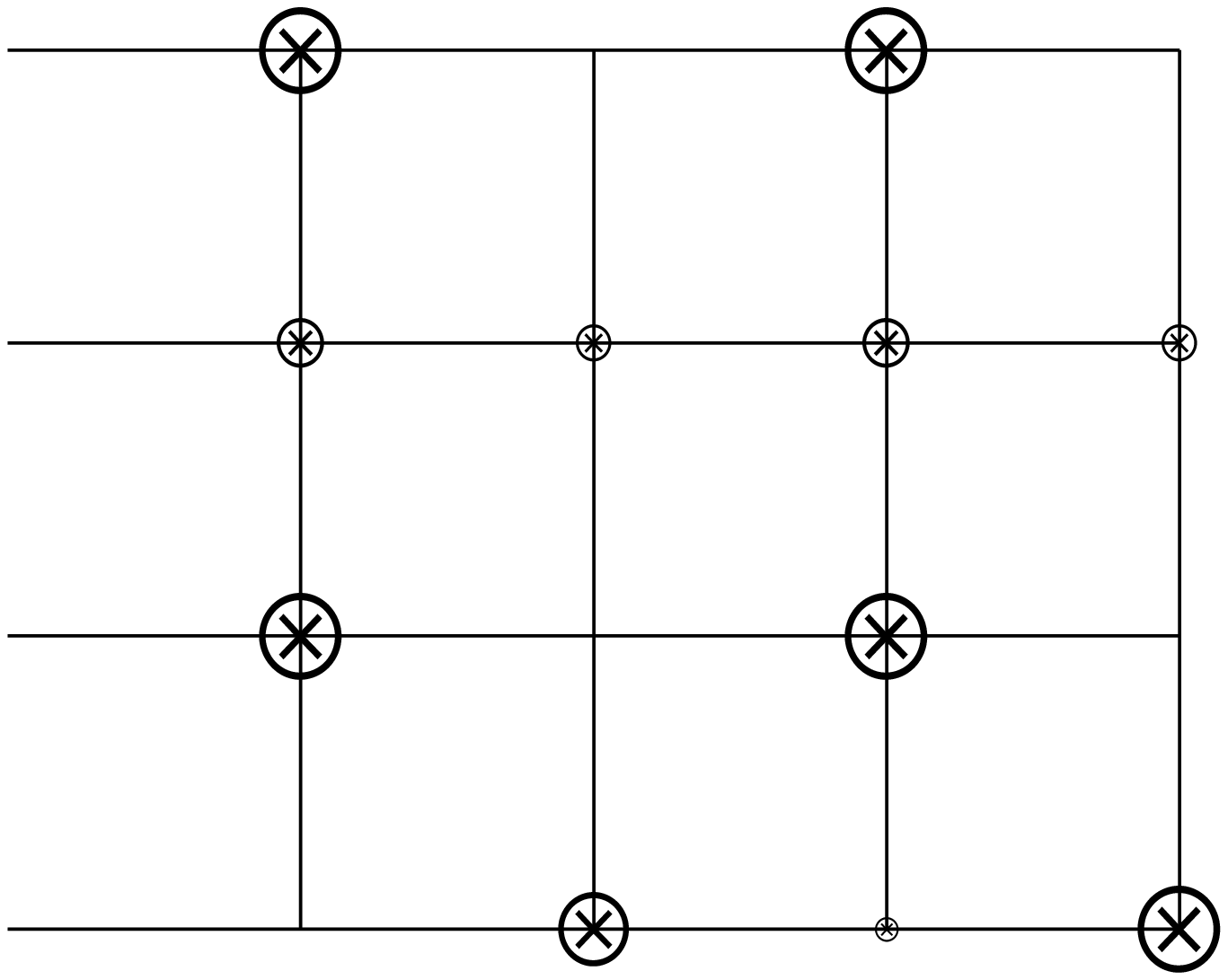}}
\caption {The local current (left column) and density (right column) for
the several realizations at $U=10V$.  The magnitude of the density at a 
particular site is indicated by the size of the circle
\label{fig.2}}
\end{figure}

Indeed for interaction strength corresponding to $r_s>1$ the system described
by the Hamiltonian in Eq. (\ref{hamil}) is known to exhibit short range 
density correlations which develop into a CDW (charge density wave) for
stronger interactions \cite{ba}. The charge density and the persistent current
for some typical realization are shown in Fig. \ref{fig.2}. Most of
the current flows in channels which correspond to
dislocations in the pinned charge density wave clearly seen in the density 
plot of the different realizations. This is a feature which is characteristic 
of plastic flows.

\begin{figure}
\centerline{\epsfxsize = 4in \epsffile{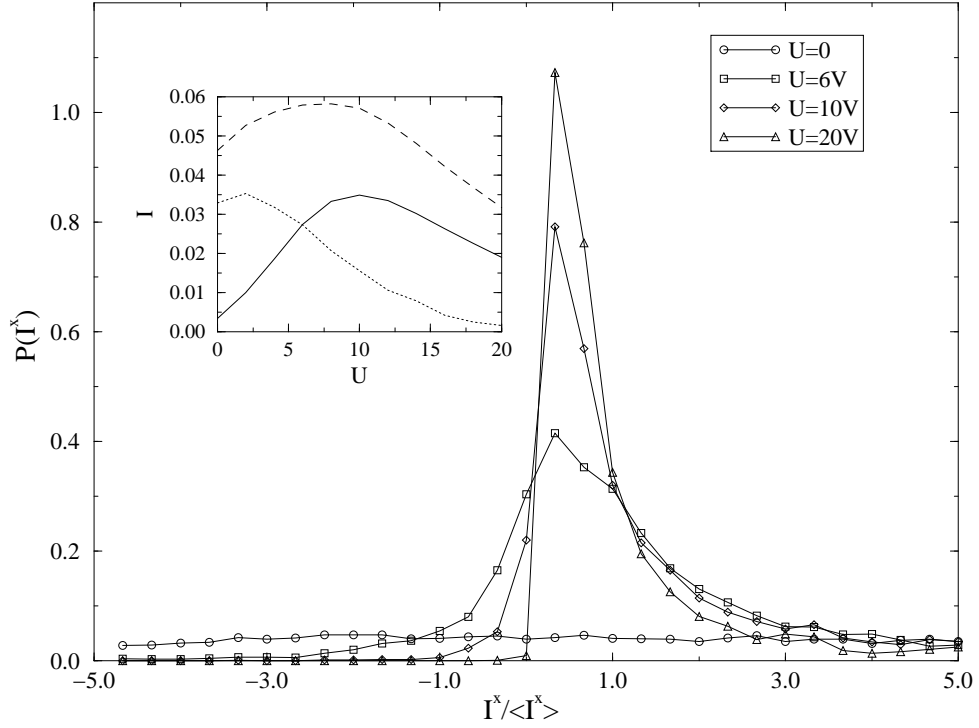}}
\caption {The distribution of the bond current in the $\hat x$ direction
for different values of $U$. Inset: average (full line) and typical
(dashed line) bond currents in the $\hat x$ direction and 
typical (dotted line) bond currents in the $\hat y$ direction in units of 
the hopping matrix $V$.
\label{fig.3}}
\end{figure}

A more quantitative measure for the change that the persistent current in
the sample undergoes due to the e-e interactions is given by 
the correlation between the currents at different locations. This correlation
may be formulated in the following way:
\begin{eqnarray}
C(r,\Phi) = 
{{\sum_{k,j} \langle I_{k,j}^{x}(\Phi) I_{k,j+r}^{x}(\Phi) \rangle 
- \langle I^{x} \rangle^2} \over
{\langle (I^{x}(\Phi))^2 \rangle  - \langle I^{x}(\Phi) \rangle^2}}
\label{cor}
\end{eqnarray} 
where
\begin{eqnarray}
\langle (I^{a}(\Phi))^N \rangle
= \langle (I_{k,j}^{a}(\Phi))^N \rangle 
\label{iav}
\end{eqnarray} 
and $\langle \ldots \rangle$ denotes average over different realizations
of disorder. The average current $\langle I^{a}(\Phi) \rangle$ 
as well as the typical current $\sqrt{\langle (I^{a}(\Phi))^2 \rangle}$
on a bond 
are plotted in the inset of Fig. \ref{fig.3}, while the current per bond
distribution $P(I^{x})$ for different values of interaction strength
are plotted in Fig. \ref{fig.3}. 
As in the case of the
total average current shown in Ref. \cite{pc} the average
current in the $\hat x$ direction increases
up to $U=10V$ and then decreases, while
the typical current is 
somewhat enhanced up to $U=10V$ and then decreases.
Of course the average current in the
$\hat y$ direction is zero and the typical current is 
somewhat enhanced in the weak interaction
regime ($r_s<1$ corresponding to $U<\sqrt{32/\pi}
\sim 3$, but suppressed for higher values of interaction. 

The average current is strongly enhanced in the regime of $r_s<3$.
It is clear from the typical bond current dependence on $U$ and from the
distribution that the enhancement is not due
to an enhancement of the typical current in a certain realization but due to
the fact that in the interacting case the local current in the $\hat x$
direction for almost all realizations
is paramagnetic while in the non-interacting case there is an almost equal 
probability of the current being para or diamagnetic. 

\begin{figure}
\centerline{\epsfxsize = 4in \epsffile{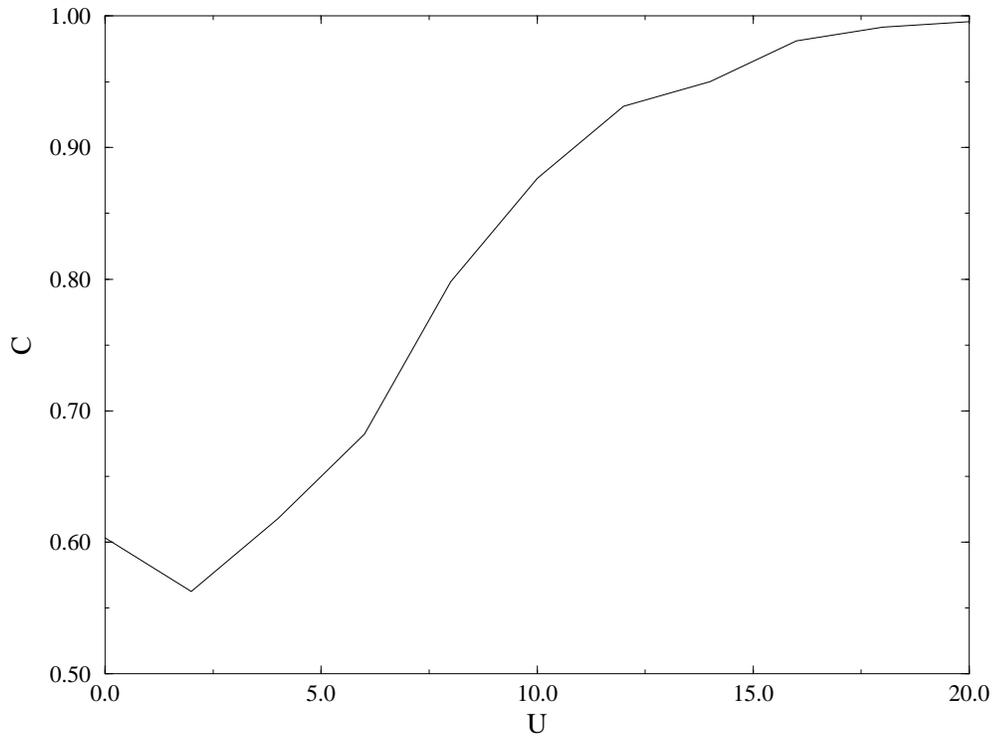}}
\caption {The bond current correlation in the $\hat x$ direction
as function of the interaction
strength $U$.
\label{fig.4}}
\end{figure}

The correlation between the persistent currents at neighboring 
bonds as function of the interaction strength is
seen in the plot of $C(r=1,\Phi=\pi/2)$ presented in Fig. \ref{fig.4}. 
While in the weak interaction RPA regime there is no
enhancement of the local current correlations due to the increase in the
interaction strength,
there is a strong enhancement of these correlations in the strong interaction
low density regime $r_s>1$. Thus, the impression one gets from the current
plots shown in Figs. \ref{fig.1} and \ref{fig.2} of a transition in the
current characteristics of the sample, is confirmed.

The reason for the enhancement of the average total current 
becomes now quite clear. 
While in the weak interaction RPA regime there is some 
enhancement in the average current due to interactions, 
this enhancement is
not related to correlations in the local current. On the other hand, in the
strongly interacting low density regime the enhancement is connected to an
ordering of the local persistent current.
The local persistent current in the strongly interacting
case must flow along dislocation lines. There is only a small probability
of forming closed loops and changing directions resulting in the fact that
the current in most realizations flows in the same direction and the
distribution of the current is almost exclusively paramagnetic.

In conclusion, strong interactions (or low densities)
impose a significant modification in the nature of the local persistent current
in a 2D sample. The current is transformed from a diffusive current to a 
plastic flow along dislocations of the emerging pinned CDW of the system.
This transformation causes a strong enhancement in the total persistent
current of the system since there is no suppression of the average 
current due to the fact that in different realizations current might flow 
in arbitrary directions.

We are grateful to the Israeli Academy of Sciences and
Humanities research center ``Strongly Interacting Electrons in Restricted
Geometries '' for financial support.

\end{document}